\documentclass[showpacs,pra,twocolumn,aps]{revtex4}
\usepackage{graphicx}
\usepackage{dcolumn}
\usepackage{bm}

\newcommand{\ket}[1]{|#1\rangle}

\providecommand{\openone}{\leavevmode\hbox{\small1\kern-3.8pt\normalsize1}}

\begin{document}

\title{Harnessing non-Markovian quantum memory by environmental coupling}

\author{Zhong-Xiao Man,$^{1,5}$ Yun-Jie Xia,$^{1}$ and Rosario Lo Franco$^{2,3,4}$}
\affiliation{$^1$Shandong Provincial Key Laboratory of Laser Polarization
and Information Technology, Department of Physics, Qufu Normal University, Qufu 273165, China\\
$^2$Dipartimento di Fisica e Chimica, Universit\`{a} di Palermo, via Archirafi 36, 90123 Palermo, Italy\\
$^3$Instituto de F{\'{i}}sica de S{\~{a}}o Carlos, Universidade de S{\~{a}}o Paulo, CP 369, 13560-970 S{\~{a}}o Carlos, SP, Brasil\\
$^4$School of Mathematical Sciences, The University of Nottingham, University Park, Nottingham NG7 2RD, United Kingdom\\
$^5$Key Laboratory of Quantum Information, University of Science and Technology of China, Chinese Academy of Sciences, Hefei 230026,  China}

\begin{abstract}
Controlling the non-Markovian dynamics of open quantum systems is essential in quantum information technology since it plays a crucial role in preserving quantum memory. Albeit in many realistic scenarios the quantum system can simultaneously interact with composite environments, this condition remains little understood, particularly regarding the effect of the coupling between environmental parts.  
We analyze the non-Markovian behavior of a qubit interacting at the same time with two coupled single-mode cavities which in turn dissipate into memoryless or memory-keeping reservoirs. We show that increasing the control parameter, that is the two-mode coupling, allows for triggering and enhancing a non-Markovian dynamics for the qubit starting from a Markovian one in absence of coupling. Surprisingly, if the qubit dynamics is non-Markovian for zero control parameter, increasing the latter enables multiple transitions from non-Markovian to Markovian regimes.  
These results hold independently on the nature of the reservoirs. This work highlights that suitably engineering the coupling between parts of a compound environment can efficiently harness the quantum memory, stored in a qubit, based on non-Markovianity.   
\end{abstract}


\maketitle

\section{Introduction}

A thorough understanding of the dynamics of an open quantum system has experienced a long term
pursuit \cite{open} and nowadays attracts ever-increasing attention due to the development of
quantum information technology \cite{infor,ladd} that employs open quantum systems as basic resource.
In the theory of open quantum systems, the non-Markovian dynamics is one of main concerns being linked to the preservation of quantum memory stored in a quantum system \cite{open,carusoreview,lofrancoreview}. It arises in many realistic situations 
\cite{carusoreview,lofrancoreview,RivasReview,NonM1,NonM2,NonM3,NonM4,NonM5,NonM6,NonM7,NonM8,NonM9,NonM10,NonM11,NonM12,NonM13,NonM14,NonM15,NonM16,NonM17}, 
and also proves useful in quantum information processing such as quantum state engineering, quantum channel capacity and quantum
control \cite{carusoreview,BB,darrigo,orieux,lofrancoPRB,sabrinaSciRep}. The degree of a non-Markovian evolution, the so-called non-Markovianity,
can be quantified by different measures based on dynamical features of the system capable to grasp the memory effects of the environment on the system evolution \cite{sabrinaSciRep,BLP,LPP,RHP,LFS,wolf,Lu,dariuszsabrina}.
So far, many factors that can trigger and modify the non-Markovian dynamics have been
found, as strong system-environment coupling, structured
reservoirs, low temperatures, and initial system-environment correlations
\cite{open,carusoreview,lofrancoreview,RivasReview,ini-corr1,ini-corr2, ini-corr3,ini-corr4,ini-corr5}. Apart from these mechanisms, 
some other peculiar conditions such as classical environments \cite{class,class-exp} and
environmental initial correlations \cite{nonlocal,nonlocal-exp} have also been predicted
and experimentally demonstrated \cite{class-exp,nonlocal-exp} enabling emergence of non-Markovianity.

In the conventional study, one usually considers the quantum system being coupled to
a single environment \cite{open,carusoreview,lofrancoreview}. However, in several realistic scenarios the system may be
simultaneously influenced by many environments \cite{dot,NV,sili,interf,compe}. For
instance, in a quantum dot the electron spin may be affected strongly by
the surrounding nuclei and weakly by the phonons \cite{dot}. The neighbor
nitrogen impurities constitute the principal bath for a nitrogen-vacancy
center, while the carbon-13 nuclear spins also have some interaction with it
\cite{NV}. A similar situation also occurs for a single-donor electron spin in
silicon \cite{sili}. Motivated by these practical situations \cite{dot,NV,sili}, some efforts have been
devoted to study the effects of multiple environments on the dynamics of an open system \cite{interf,compe,MAX14,Yu14,tiered}.
Quantum interference effects have been found to occur between independent reservoirs when all of them interact with a quantum system and are in non-Markovian regimes, which qualitatively modify the dynamics of the interested system \cite{interf}.
The dynamics of a spin simultaneously coupled to two decoherence channels, one Markovian and the other non-Markovian, has been analyzed with respect to the different decoherence mechanisms \cite{compe}. As is known, a qubit (i.e., a two-level system) interacting with a single vacuum bosonic reservoir may exhibit Markovian or non-Markovian dynamics depending on the strength of the system-reservoir coupling \cite{open}.
By contrast, if the qubit simultaneously interacts with several reservoirs, its dynamics can be always non-Markovian provided that the number of the contributing reservoirs is greater than a critical value \cite{MAX14}. The dynamics of a qubit coupled to a hierarchical environment made of a
single-mode cavity and a structured reservoir with Lorentzian spectral density has been studied, showing that a shorter (longer) memory time of
the reservoir does not generally mean a smaller (larger) non-Markovianity of the system \cite{Yu14}.
A new analytical method based on a phase space representation of the system density matrix has been also proposed to study the dynamics of a discrete system in a two-tiered non-Markovian environment \cite{tiered}.

In the treatments of composite environments \cite{interf,compe,MAX14}, the role of the coupling between environmental parts is not typically taken into account. Despite this, the knowledge of how this environmental parameter influences the non-Markovian character of an open quantum system would provide insightful developments for engineering and controlling quantum memories for possible technological applications. Therefore, this aspect deserves careful investigation, possibly starting from a paradigmatic model where it can simply emerge and be understood. 
Here we choose a model which complies with this requirement, namely a qubit interacting at the same time with two coupled single-mode cavities which in turn dissipate photons into their own memoryless (Markovian) or memory-keeping (non-Markovian) reservoirs. This system finds its natural implementation in nowadays technologies of circuit quantum electrodynamics \cite{circuitQED} and also in simulating all-optical setups \cite{chiuriSciRep}.
We shall show that the coupling strength between the two modes can harness the qubit non-Markovianity in different and even counterintuitive ways, independently of the nature of the reservoirs. The paper is organized as follows. In Sec.~\ref{system} we describe the system of interest. In Sec.~\ref{memorylessR} we study the case of memoryless reservoirs, while in Sec.~\ref{memoryR} we investigate the case of memory-keeping reservoirs. In Sec.~\ref{conclusion} we summarize our conclusions.

\section{The system}\label{system}

\begin{figure}[tbp]
\begin{center}
\includegraphics[width=0.46\textwidth]{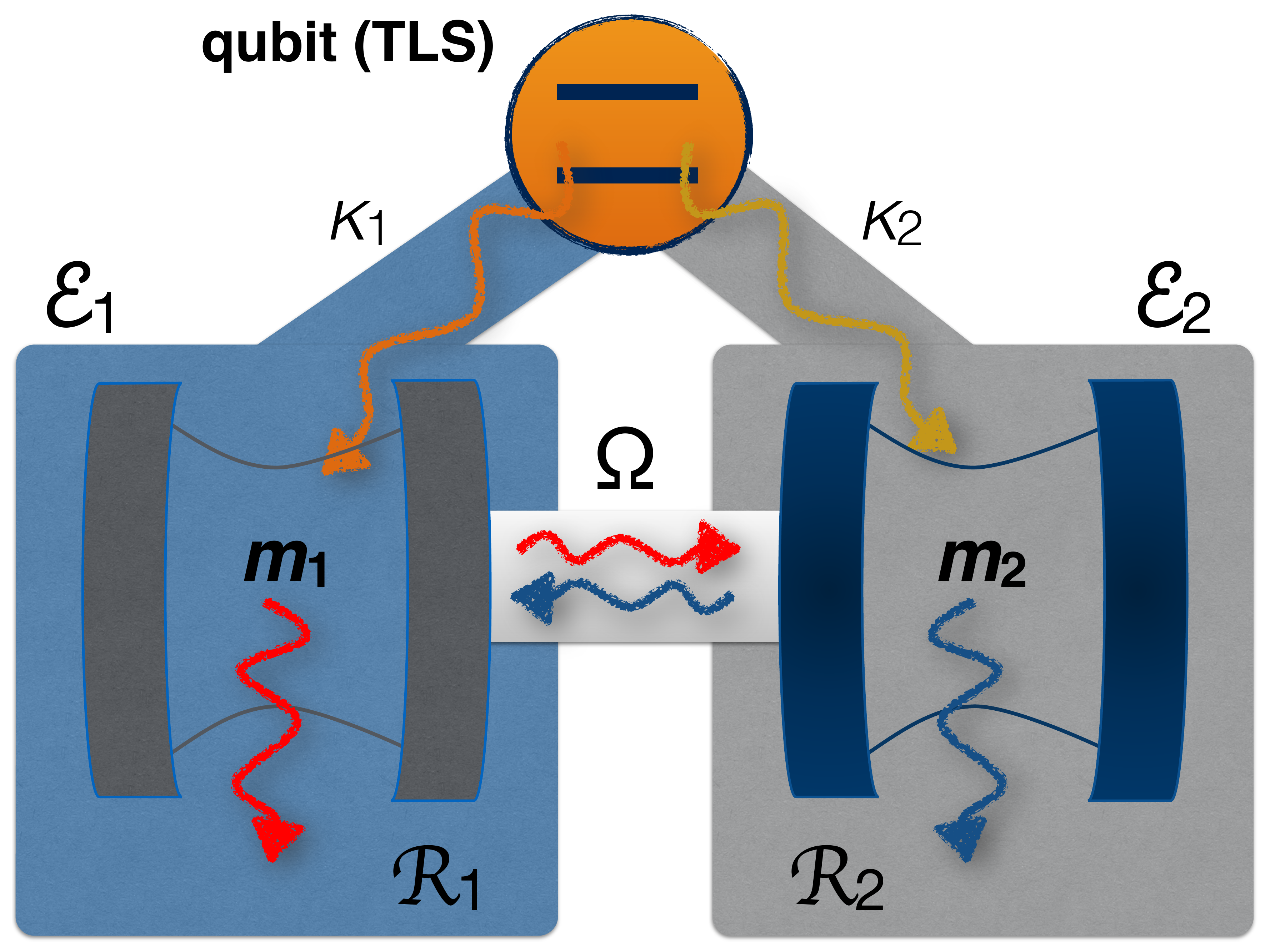}
\end{center}
\caption{\textbf{Pictorial description of the global system.} A qubit, i.e. a two-level system (TLS), simultaneously interacts with two environments 
$\mathcal{E}_n$ ($n=1,2$), each containing a single cavity mode $m_n$ that in turns is affected by a vacuum (zero temperature) reservoir 
$\mathcal{R}_n$. The qubit is directly coupled to each mode $m_n$ with strength $\kappa_n$. The two cavity modes $m_1$, $m_2$ are coupled with strength $\Omega$. }
\label{fig:system}
\end{figure}

Our global system consists of a qubit $s$ simultaneously interacting with two environments $\mathcal{E}_{1}$
and $\mathcal{E}_{2}$. To fix the ideas, we model each environment $\mathcal{E}_{n}$ ($n=1,2$)
as a bosonic mode $m_{n}$ decaying to a zero-temperature bosonic reservoir $\mathcal{R}_{n}$, as depicted in Fig.~\ref{fig:system}. 
The qubit is here meant as the quantum memory whose efficiency is to be quantified by its non-Markovianity, that is by the degree of non-Markovian evolution. The interaction of the two environments is due to the coupling of the two bosonic cavity modes, which instead plays the role of a control parameter for the non-Markovianity of the qubit.
The coupling strength of the qubit with each mode $m_n$ is $\kappa_{n}$, while $\Omega$ denotes 
the coupling between the two modes. For the sake of simplicity, we assume that the two modes 
have the same frequency $\omega_c$ which in turn is equal to the qubit transition frequency $\omega_0$, 
that is $\omega _{0}=\omega _{c}$. 

The total Hamiltonian is given by ($\hbar = 1$)
\begin{eqnarray}\label{H1}
\hat{H}&=&\hat{H}_s+\sum_{n=1}^{2}[\hat{H}_{m_n}+\hat{H}_{\mathcal{R}_n}+\hat{H}_{sm_n}+\hat{H}_{m_n\mathcal{R}_n}]\nonumber\\
&&+\hat{H}_{m_1m_2},
\end{eqnarray}
where $\hat{H}_s=(\omega _{0}/2)\hat{\sigma}_{z}$ is the qubit Hamiltonian, $\hat{H}_{m_n}=\omega_{c}\hat{a}_{n}^{\dag }\hat{a}_{n}$ the mode Hamiltonian, $\hat{H}_{\mathcal{R}_n}=\sum_{k}\omega_{n,k}\hat{b}^{\dag}_{n,k}\hat{b}_{n,k}$ the reservoir Hamiltonian, 
$\hat{H}_{sm_n}= \kappa_{n}(\hat{a}_{n}^{\dag}\hat{\sigma} _{-}+\hat{a}_{n}\hat{\sigma} _{+})$ the qubit-mode interaction Hamiltonian, 
$\hat{H}_{m_n\mathcal{R}_n}=\sum_{k}g_{n,k}(\hat{a}_{n}\hat{b}_{n,k}^{\dag}+\hat{a}_{n}^{\dag}\hat{b}_{n,k})$ the mode-reservoir interaction Hamiltonian
and $\hat{H}_{m_1m_2}=\Omega(\hat{a}_{1}^{\dag}\hat{a}_{2}+\hat{a}_{1}\hat{a}_{2}^{\dag})$ the interaction Hamiltonian between the two modes.
In the expressions above $\hat{\sigma}_{z}=\left|1\right\rangle\left\langle1\right|-\left|0\right\rangle\left\langle0\right|$
is a Pauli operator for the system with transition frequency $\omega _{0}$, $\hat{\sigma} _{\pm }$\ represent
the raising and lowering operators of the qubit, $\hat{a}_{n}$\ $(\hat{a}_{n}^{\dag })$
the annihilation (creation) operator of mode $m_{n}$. Furthermore, in the Hamiltonians involving the reservoirs 
$\hat{b}_{n,k}$\ $(\hat{b}_{n,k}^{\dag })$ is the annihilation (creation) operator of field mode $k$\ with frequency $\omega _{n,k}$\ of
reservoir $\mathcal{R}_{n},$ and $g_{n,k}$ denotes the coupling
of the mode $m_{n}$ with the mode $k$ of its own reservoir $\mathcal{R}_{n}$.
In the interaction picture, the total Hamiltonian can be expressed as
\begin{eqnarray}\label{H2}
\hat{H}_\mathrm{int}&=&\sum_{n=1}^{2}\hat{H}_{sm_n}+\hat{H}_{m_1m_2}\nonumber\\
&+&\sum_{n=1}^{2}\sum_{k}g_{n,k}(\hat{a}_{n}\hat{b}_{n,k}^{\dag}e^{i\Delta_{n,k}t}+\hat{a}_{n}^{\dag}\hat{b}_{n,k}e^{-i\Delta_{n,k}t}),
\end{eqnarray}
where $\Delta_{n,k}=\omega_{n,k}-\omega_{0}$.

The reservoirs $\mathcal{R}_{n}$ of the global system can be either memoryless (Markovian) or memory-keeping (non-Markovian). Depending on the kind of reservoir, different methods are used to obtain the reduced dynamics of the qubit. In the following two sections we study these two cases.

\section{Memoryless reservoirs}\label{memorylessR}
In this section we consider both the reservoirs $\mathcal{R}_n$ as vacuum Markovian ones, their correlation times being much smaller than the single-mode relaxation times. Although our system can be exactly solved (see Sec.~\ref{memoryR}), we first treat it under the Markov approximation 
since this analysis constitutes a strategical first step in order to strongly evidence the crucial role of the two-mode coupling parameter to harness quantum non-Markovianity for the dynamics of the qubit even under this condition.
In this case, the density operator $\rho (t)$ of the qubit plus the two modes obeys the following master equation \cite{open}
\begin{eqnarray} \label{ro}
\dot{\rho}(t) &=&-i[\hat{H},\rho(t)]\nonumber\\
&-&\sum_{n=1}^{2}\frac{\Gamma _{n}}{2}[a_{n}^{\dag }a_{n}\rho(t)-2a_{n}\rho(t)a_{n}^{\dag }+\rho(t)a_{n}^{\dag }a_{n}], 
\end{eqnarray}
where $\dot{\rho}(t)\equiv d\rho(t)/dt$, $\hat{H}$ is given by Eq.~(\ref{H1}) without the terms involving the reservoirs and $\Gamma _{n}$ denotes the decay rate of the mode $m_{n}$.
We initially take the qubit in its excited state $\left| 1\right\rangle_{s}$ and both modes in the
ground states $\left| 00\right\rangle _{m_{1},m_{2}}$, so that the initial overall state is $\rho(0)=\left| 100\right\rangle\left\langle 100\right|$, where the first, second and third element correspond to the qubit $s$, mode $m_1$ and mode $m_2$, respectively. 
Since there exist at most one excitation in the total system at any time, we can make the ansatz for $\rho(t)$ in the form
\begin{equation}\label{rot}
\rho (t)=\left( 1-\lambda (t)\right) \left| \psi (t)\right\rangle\left\langle \psi (t)\right| 
+\lambda (t)\left| 000\right\rangle\left\langle 000\right|,
\end{equation}
where $0\leq \lambda (t)\leq 1$\ with $\lambda (0)=0$ and 
$\left| \psi(t)\right\rangle=h(t)\left| 100\right\rangle +c_{1}(t)\left|010\right\rangle 
+c_{2}(t)\left| 001\right\rangle$
with $h(0)=1$\ and $c_{1}(0)=c_{2}(0)=0.$ It is convenient to
introduce the unnormalized state vector \cite{pseu1}
\begin{eqnarray}  \label{unnor}
\left| \widetilde{\psi }(t)\right\rangle &\equiv &\sqrt{1-\lambda (t)}\left| \psi (t)\right\rangle \nonumber \\
&=&\widetilde{h}(t)\left| 100\right\rangle +\widetilde{c}
_{1}(t)\left| 010\right\rangle+\widetilde{c}_{2}(t)\left|001\right\rangle,
\end{eqnarray}
where $\widetilde{h}(t)\equiv \sqrt{1-\lambda (t)}h(t)$ represents the
probability amplitude of the qubit and $\widetilde{c}_{n}(t)\equiv \sqrt{1-\lambda (t)}c_{n}(t)$ 
that of the mode $m_n$ being in their excited
states. In terms of the unnormalized state vector we then get
\begin{equation} \label{ronga}
\rho (t)=| \widetilde{\psi }(t)\rangle\langle
\widetilde{\psi }(t)| +\lambda (t)| 000\rangle\langle 000|. 
\end{equation}
Inserting this expression in Eq.~(\ref{ro}), the time-dependent amplitudes $\widetilde{h}(t),$\ $\widetilde{c}_{1}(t),$\ $\widetilde{c}_{2}(t)$ of Eq. (\ref{unnor}) are determined by a set of differential equations as
\begin{eqnarray}
i\frac{d\widetilde{h}(t)}{dt} &=&\omega _{0}\widetilde{h}(t)+
\kappa_{1}\widetilde{c}_{1}(t)+\kappa_{2}\widetilde{c}_{2}(t), \nonumber \\
i\frac{d\widetilde{c}_{1}(t)}{dt} &=&\left( \omega _{c}-\frac{i}{2}\Gamma
_{1}\right) \widetilde{c}_{1}(t)+\kappa _{1}\widetilde{h}(t)+
\Omega \widetilde{c}_{2}(t),  \nonumber \\
i\frac{d\widetilde{c}_{2}(t)}{dt} &=&\left( \omega _{c}-\frac{i}{2}\Gamma
_{2}\right) \widetilde{c}_{2}(t)+\kappa _{2}\widetilde{h}(t)+\Omega \widetilde{c}_{1}(t).
\label{eqs}
\end{eqnarray}
The above differential equations can be solved by means of standard Laplace
transformations combined with numerical simulations to obtain the reduced
density operators of the qubit as well as of each of the modes.

\begin{figure}[tbp]
\begin{center}
{\includegraphics[width=3.5in]{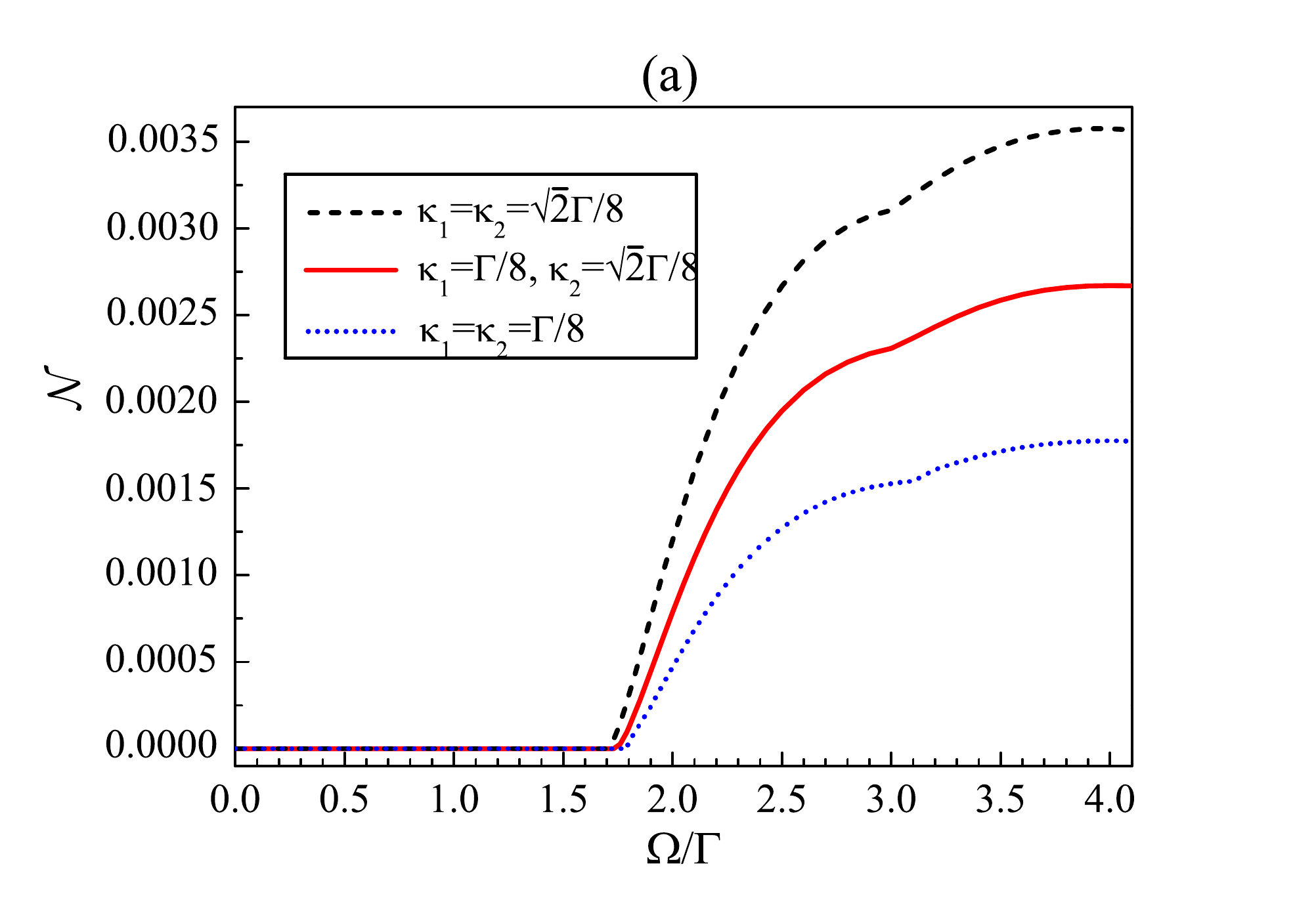}
\includegraphics[width=3.5in]{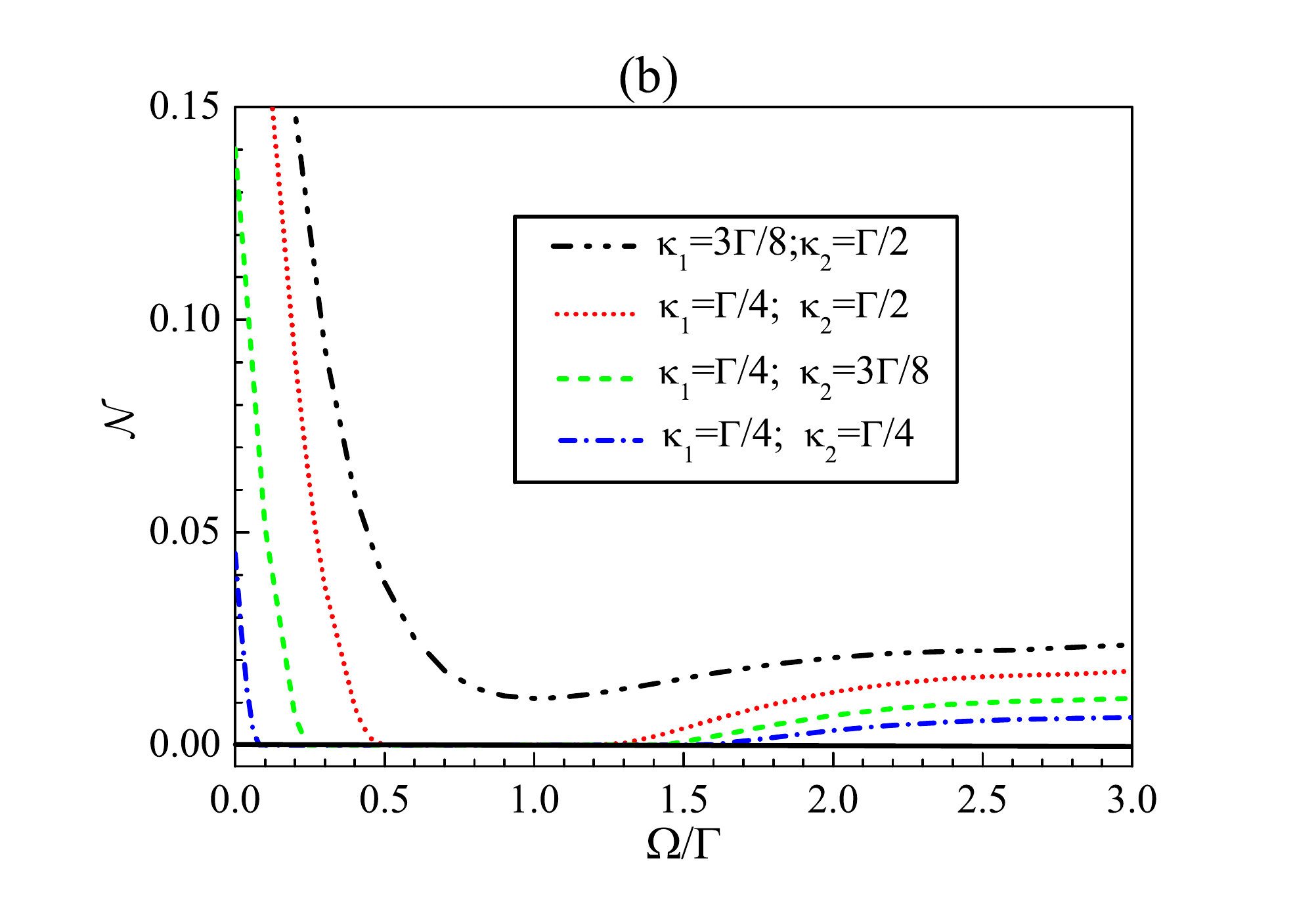}}
\end{center}
\caption{Non-Markovianity measure $\mathcal{N}$ as a function of the coupling
constant $\Omega /\Gamma$ between the two modes for (a) weak and (b) strong
system-mode coupling regimes.}
\label{NM}
\end{figure}

To quantify the non-Markovianity we adopt a measure based on the dynamics of the trace distance between two different initial states
$\rho_{1}(0)$ and $\rho _{2}(0)$ of an open system \cite{BLP}.
A Markovian evolution can never increase the distinguishability between different initial
states in terms of their trace distance, hence a nonmonotonic time behavior
of the latter would signify non-Markovian dynamics of the system.
Such a measure is consistent with the interpretation of non-Markovianity in terms of a backflow of information from the environment to the system, which is responsible for the distance (state-distinguishability) growth.
Based on this concept, the non-Markovianity can be quantified by a measure $\mathcal{N}$ defined as \cite{BLP}
\begin{equation}
\mathcal{N}=\max_{\rho _{1}(0),\rho _{2}(0)}\int_{\sigma >0}\sigma [t,\rho
_{1}(0),\rho _{2}(0)]dt,  \label{N}
\end{equation}
in which $\sigma [t,\rho _{1}(0),\rho _{2}(0)]=dD[\rho _{1}(t),\rho
_{2}(t)]/dt$ is the rate of change of the trace distance given by
\begin{equation}
D[\rho _{1}(t),\rho _{2}(t)]=\frac{1}{2}\mathrm{Tr}|\rho _{1}(t)-\rho
_{2}(t)|,  \label{Tra-Dis}
\end{equation}
where $|A|=\sqrt{A^{\dag }A}.$ In order to evaluate $\mathcal{N,}$ we have to find a specific pair of optimal initial states to
maximize the time derivative of the trace distance. In Ref. \cite{optimal},
it is proved that the pair of optimal states is associated with two
antipodal pure states on the surface of the Bloch sphere. We thus adopt a
pair of initial states $\rho _{1,2}(0)=\left| \psi _{1,2}(0)\right\rangle
\left\langle \psi _{1,2}(0)\right| $\ with $\left| \psi
_{1,2}(0)\right\rangle =(\left| 0\right\rangle \pm \left| 1\right\rangle )\sqrt{2}$ as the optimal ones throughout the paper. This allows us
to obtain the time derivative of the trace distance in the simple form
$\sigma [t,\rho _{1}(0),\rho _{2}(0)]=d|\widetilde{h}(t)|/dt$.

In the absence of the coupling between $m_1$ and $m_2$, the qubit exhibits
Markovian dynamics when the couplings of the qubit with the two modes
in terms of $\kappa_{1}/\Gamma_{1}$ and $\kappa_{2}/\Gamma_{2}$ are weak.
In this case, we show that the introduction of mode-mode coupling with sufficient strength $\Omega$
can transform the Markovian dynamics to the non-Markovian one.
In Fig.~\ref{NM}(a), we plot the non-Markovianity $\mathcal{N}$ as a function of
the scaled coupling strength $\Omega/\Gamma$ between the two modes for different values of
$\kappa_{1}$ and $\kappa_{2}$ ($\Gamma_{1}$=$\Gamma_{2}$=$\Gamma$ is assumed along the paper).
As shown in the figure, the system exhibits Markovian dynamics, individuated by $\mathcal{N}=0$, until the two modes are weakly coupled and below a certain threshold. However, when $\Omega/\Gamma$ exceeds this threshold the Markovian dynamics of the qubit changes to non-Markovian one (i.e., $\mathcal{N}>0$). In general, non-Markovianity
increases with $\Omega/\Gamma$ for the given values of $\kappa_{1}$ and $\kappa_{2}$
and is also proportional to $\kappa_{1}$, $\kappa_{2}$ for a fixed $\Omega/\Gamma$.
Therefore, the coupling of the two modes can trigger the non-Markovian dynamics of the system.

On the other hand, if the qubit-mode couplings $\kappa_{1}$, $\kappa_{2}$ are strong
the qubit exhibits non-Markovian dynamics without the need of mode-mode coupling. 
Under these conditions, how the additional coupling of the two modes influences the system non-Markovianity is to be revealed. 
From the above discussion about the case of weak qubit-mode couplings, one might expect that the mode-mode coupling would enhance the
non-Markovianity of the system. However, as shown in Fig.~\ref{NM}(b),
for different $\kappa_{1}$, $\kappa_{2}$ the relation between the
non-Markovianity $\mathcal{N}$ and the mode-mode coupling strength is nonmonotonic.
Increasing $\Omega/\Gamma$ from zero, the non-Markovianity first diminishes to a minimal value and then rises.
Remarkably, for some smaller values of $\kappa_{1}$ and $\kappa_{2}$,
the non-Markovianity can even decrease to zero (Markovian regime), remain zero for a finite range of
$\Omega/\Gamma$ and then recover nonzero values with a further increase of $\Omega/\Gamma$.
The mode-mode coupling is thus able not only to enhance the memory effects of the overall environment but also to restrain them.

\begin{figure}[tbp]
\begin{center}
\includegraphics[width=0.47\textwidth]{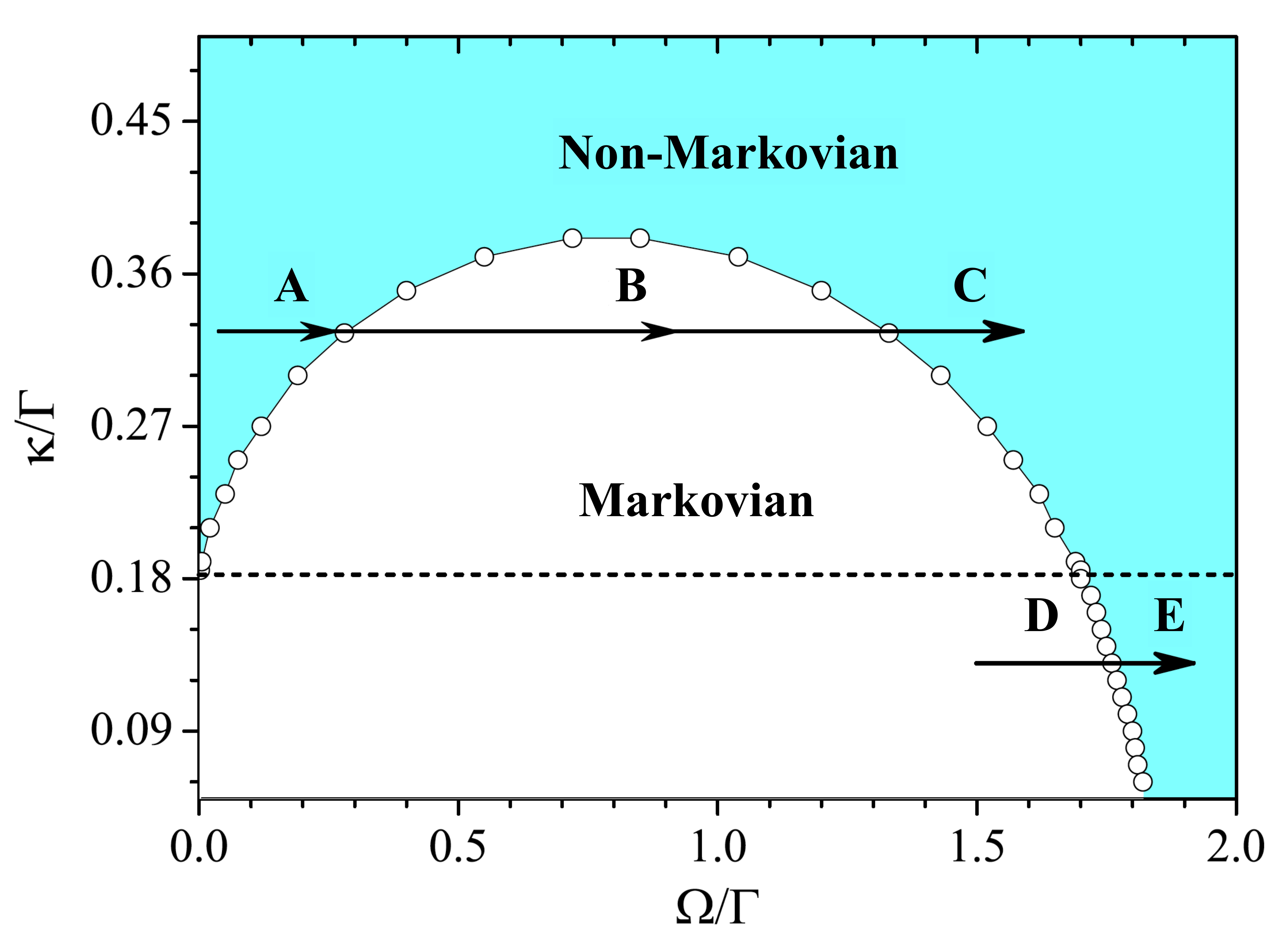}
\end{center}
\caption{Phase diagram in the $\kappa/\Gamma$-$\Omega /\Gamma $ plane for the
crossover between Markovian and non-Markovian dynamics. The colored regions represent the non-Markovian
dynamics, while the white regions denote the Markovian dynamics. The dotted line near $\kappa/\Gamma=0.18$
divides the weak and strong couplings between the system and the modes and above (below) which
is the strong (weak) system-modes coupling regime.}
\label{phase}
\end{figure}

For a comprehensive understanding of the effects of system-mode and mode-mode couplings on the qubit dynamics,
in Fig.~\ref{phase} we display the phase diagram in the $\kappa$-$\Omega$ plane ($\kappa_{1}=\kappa_{2}=\kappa$) 
of the transitions between Markovian and non-Markovian dynamics.
In the strong qubit-mode coupling regime (above the dotted line), the qubit may
experience two transitions: from non-Markovian to Markovian and again to non-Markovian
dynamics (e.g., from point $A$ to $B$ and then to $C$) with an increase of $\Omega$.
In the weak qubit-mode coupling regime (below the dotted line), an increase of the mode-mode coupling
can drive the Markovian dynamics to the non-Markovian one (e.g., from point $D$ to $E$).
Moreover, the smaller the $\kappa$, the larger the $\Omega$ required to activate the non-Markovian dynamics.

\begin{figure*}[tbp]
\begin{center}
{\includegraphics[width=0.32\textwidth]{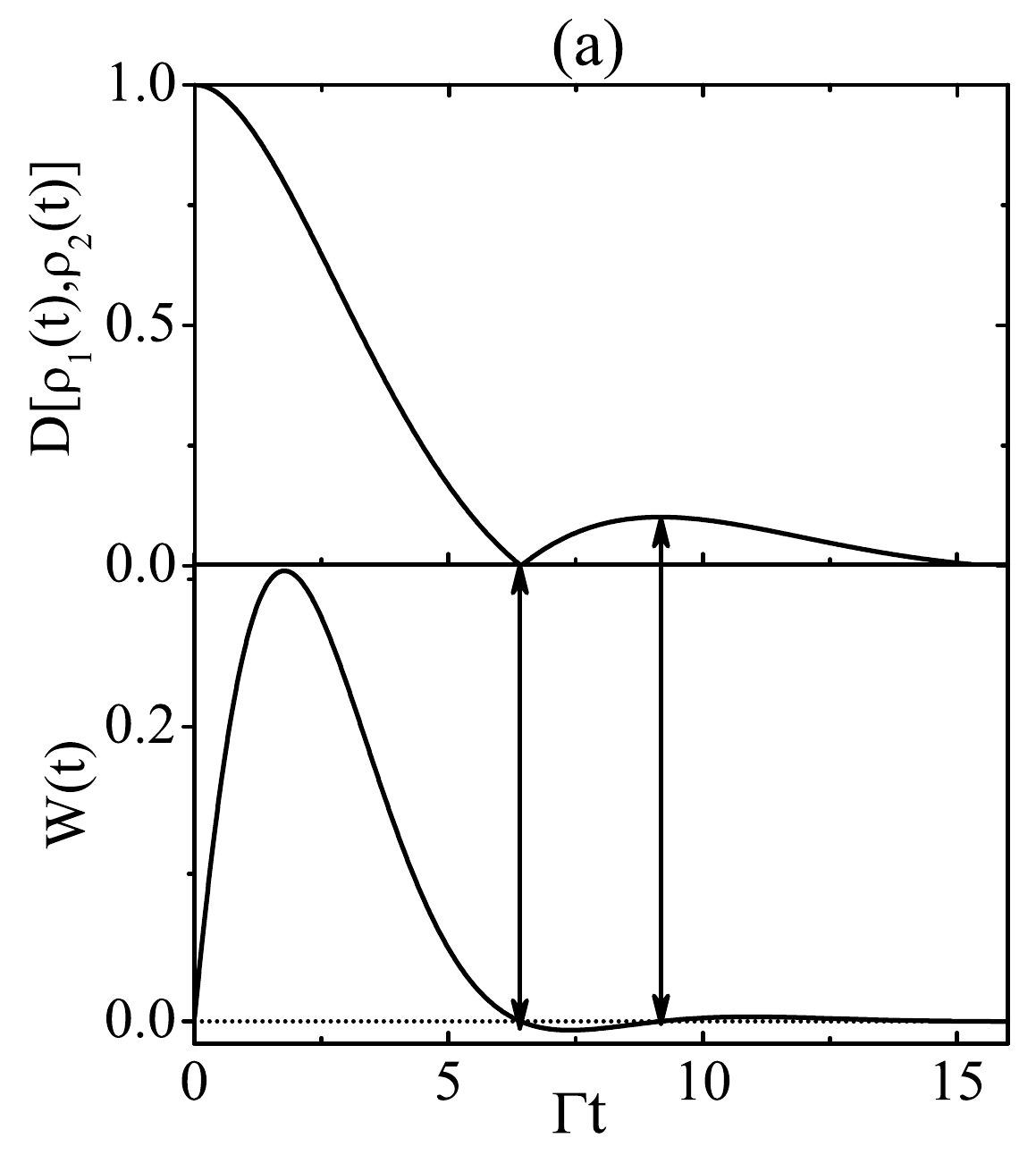}
\includegraphics[width=0.33\textwidth]{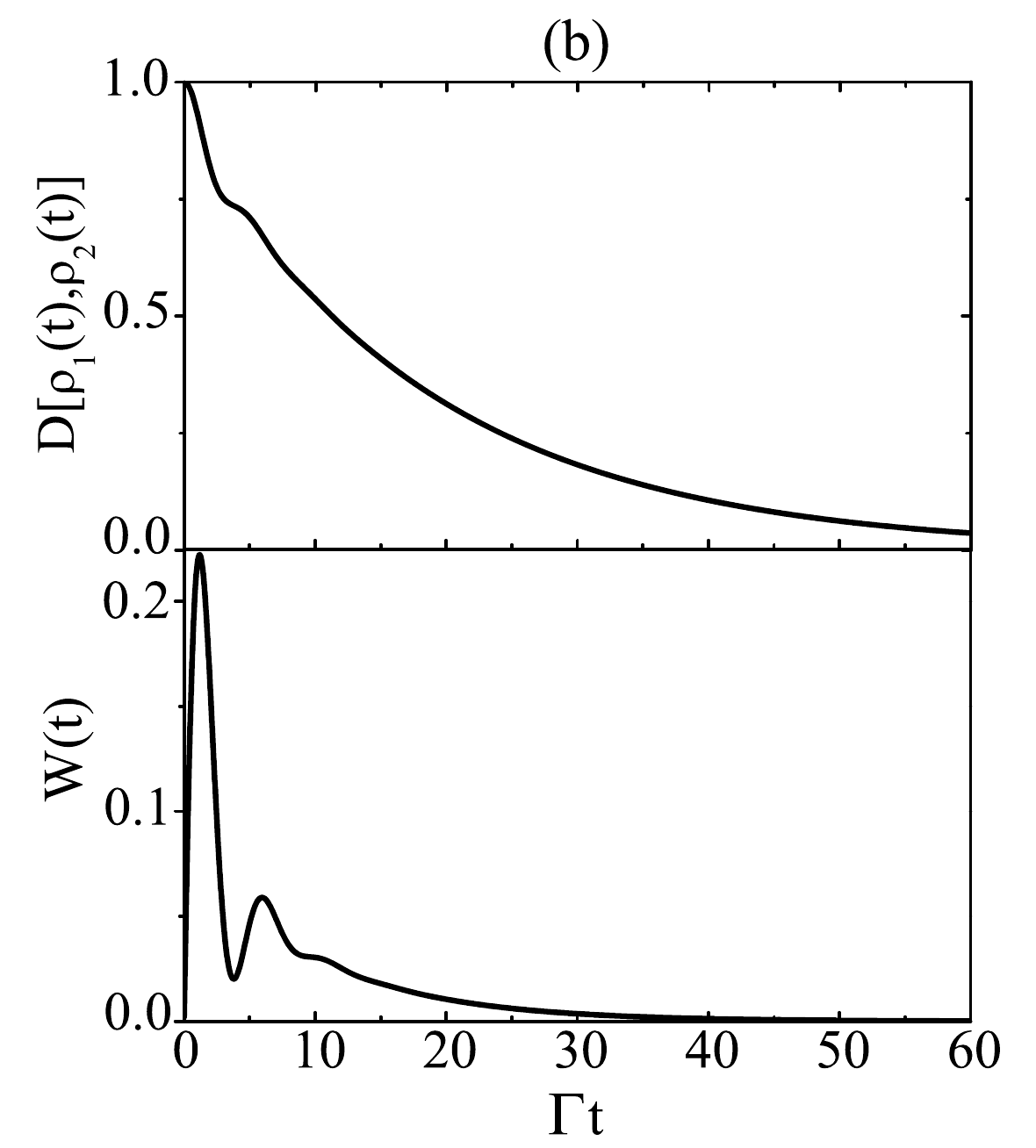}
\includegraphics[width=0.32\textwidth]{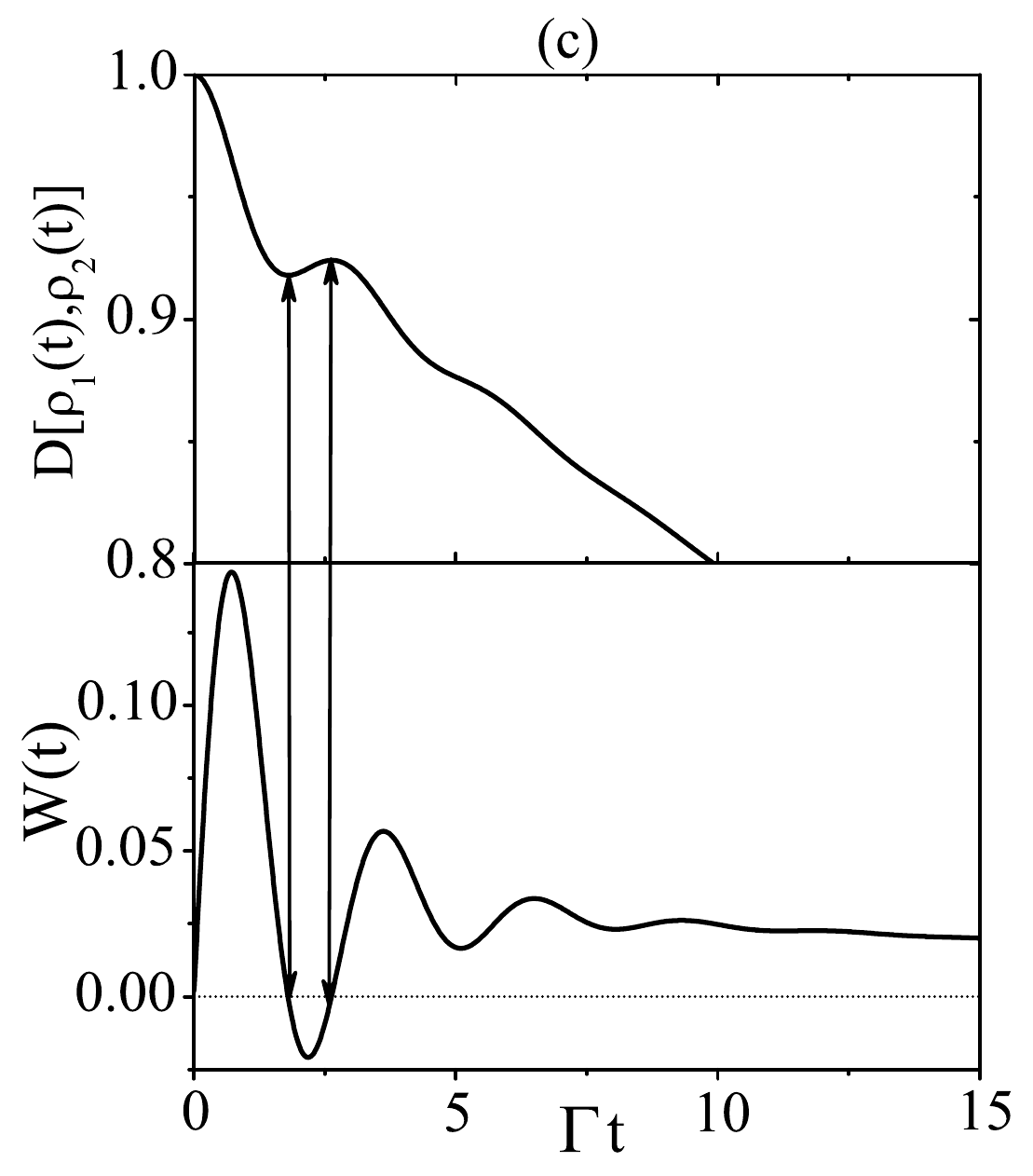}}
\end{center}
\caption{Dynamics of the trace distance $D[\rho_{1}(t),\rho_{2}(t)]$ (top plane) and the witness $W(t)$ (unit $\mathrm{s}^{-1}$) of Eq.(\ref{W}) (bottom plane), for $\kappa_{1}=\kappa_{2}=0.3 \Gamma$ and $\Omega=0$ (a), $\Omega=\Gamma$ (b) and $\Omega=2 \Gamma$ (c).}
\label{Wt}
\end{figure*}

So far, we have shown that the mode-mode coupling can trigger and modify the non-Markovianity of the
system. In fact, the two modes can be regarded as effective memories of the overall environment
since the compensated rate $W(t)$ of their population changes, given by \cite{pseu2}
\begin{equation}\label{W}
W(t)\equiv \frac{d\left(\sum_{n=1}^{2}|\widetilde{c}_{n}(t)|^{2}\right)}{dt}
+\sum_{n=1}^{2}\Gamma _{n}|\widetilde{c}_{n}(t)|^{2},
\end{equation}
completely determines the qubit non-Markovian dynamics. The meaning of Eq.~(\ref{W}) can be explained as follows.
The energy dissipations are one-way from the modes to their memoryless reservoirs so that $\Gamma _{1}$, $\Gamma _{2}$ are always positive.
If the energy of the two modes decreases (i.e., $d\left(\sum_{n=1}^{2}|\widetilde{c}_{n}(t)|^{2}\right)/dt<0$)
and this decrease is not compensated by the dissipation of the modes, quantified by $\sum_{n=1}^{2}\Gamma _{n}|\widetilde{c}_{n}(t)|^{2}$, then $W(t)<0$. This situation can only happen when part of the two-mode energy has come back to the quantum system, thus a negative value of $W(t)$ identifies a back-action (or back-flow of information) and therefore onset of non-Markovianity.
Therefore, the results discussed above indicate that the coupling between the two environmental modes, the memories, can enhance or inhibit their own memory effects on the qubit.
As a further verification, in Fig.~\ref{Wt} we compare the dynamics of the trace distance
$D[\rho_{1}(t),\rho_{2}(t)]$ of Eq.~(\ref{Tra-Dis}) and of the rate $W(t)$ to directly demonstrate that the negativity of $W(t)$ assesses qubit non-Markovianity. To this aim, we choose three points in the $\kappa$-$\Omega$ phase diagram (Fig.~\ref{phase})
with the same $\kappa=0.3\Gamma$, while $\Omega=0$, $\Gamma$ and $2\Gamma$, respectively.
These points pass through the regime transitions non-Markovian $\rightarrow$ Markovian $\rightarrow$ non-Markovian as $\Omega$ increases. As expected, $D[\rho_{1}(t),\rho_{2}(t)]$
exhibit oscillations for $\Omega=0$ and $\Omega=2\Gamma$ (see Fig.~\ref{Wt}, panels (a) and (c)),
while asymptotically decay to zero for $\Omega=\Gamma$ (Fig.~\ref{Wt}(b)).
The witness $W(t)$ becomes negative at the points where $D[\rho_{1}(t),\rho_{2}(t)]$ begins to grow and
remains negative during the whole time interval when $D[\rho_{1}(t),\rho_{2}(t)]$ increases, 
which entails an information backflow from the modes to the qubit. Differently,
$W(t)$ remains positive when $D[\rho_{1}(t),\rho_{2}(t)]$ asymptotically decays,
as in Fig.~\ref{Wt}(b). It is worth to notice that, although the qubit undergoes non-Markovian
dynamics for both $\Omega=0$ (Fig.~\ref{Wt}(a)) and $\Omega=2\Gamma$ (Fig.~\ref{Wt}(c)), the dynamical curves of the trace distance $D[\rho_{1}(t),\rho_{2}(t)]$ (state distinguishability) are very different regarding the points when it starts increasing, implying different mechanisms of information backflows in the two cases.

\section{Memory-keeping reservoirs}\label{memoryR}

In the above section, we have considered a qubit $s$ interacting with two coupling modes $m_{1}$, $m_{2}$ which are
dissipated respectively by two memoryless reservoirs $\mathcal{R}_{1}$, $\mathcal{R}_{2}$. 
Under these conditions, we have seen that the two modes are fully responsible for the memory effects of the overall environment on the qubit and their coupling can modify this effect. However, if the two modes are only components of the compound memory of the overall environment,
the way their coupling changes the overall memory effects on the qubit is to be explored. We accomplish this analysis in this section.
To this purpose, we consider a more complex situation where the coupling modes are dissipated
by structured reservoirs $\mathcal{R}_{1}$, $\mathcal{R}_{2}$ exhibiting inherent memory effects and are therefore non-Markovian \cite{open,lofrancoreview}. 

\begin{figure}[tbp]
\begin{center}
{\includegraphics[width=3.5in]{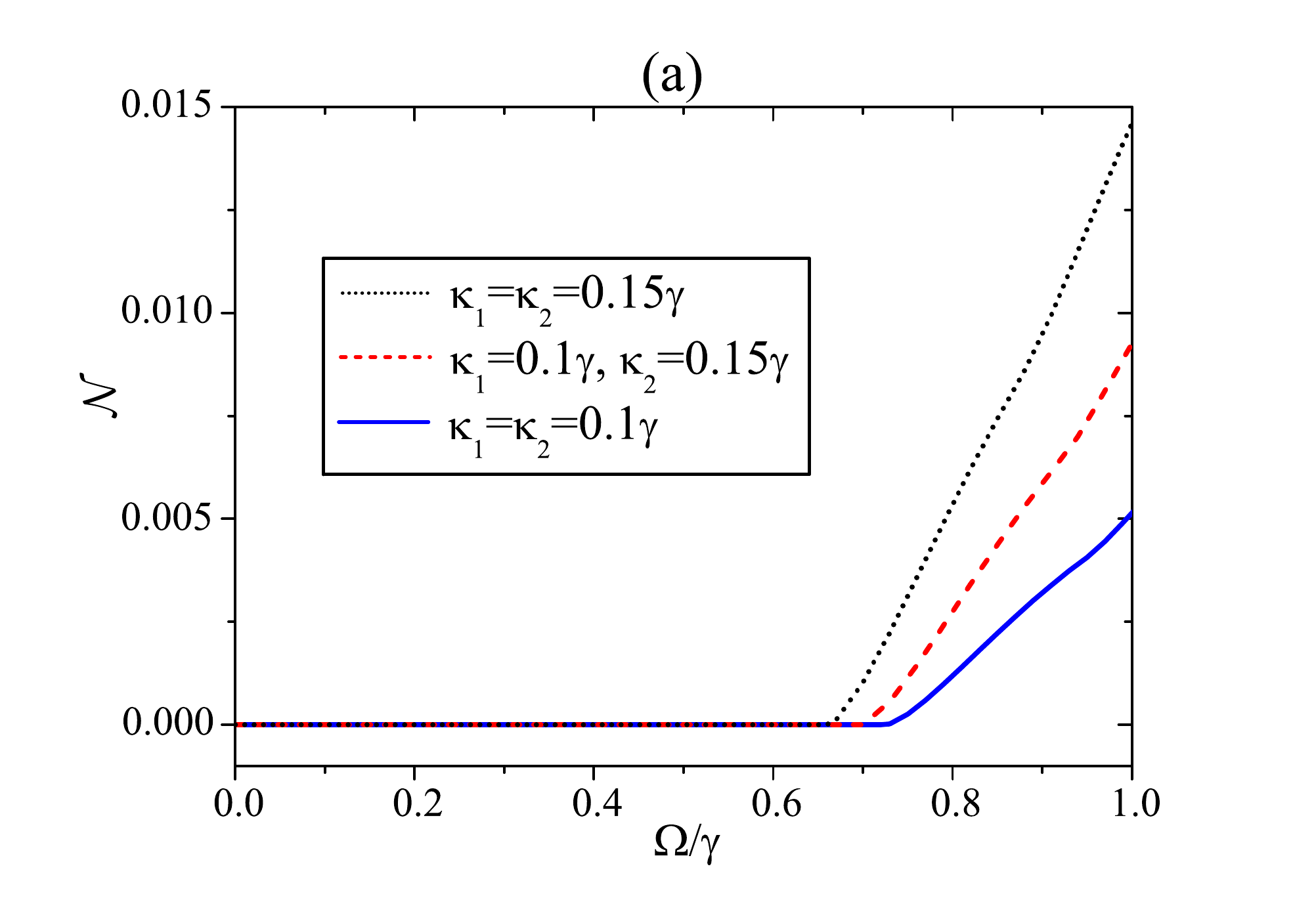}
\includegraphics[width=3.5in]{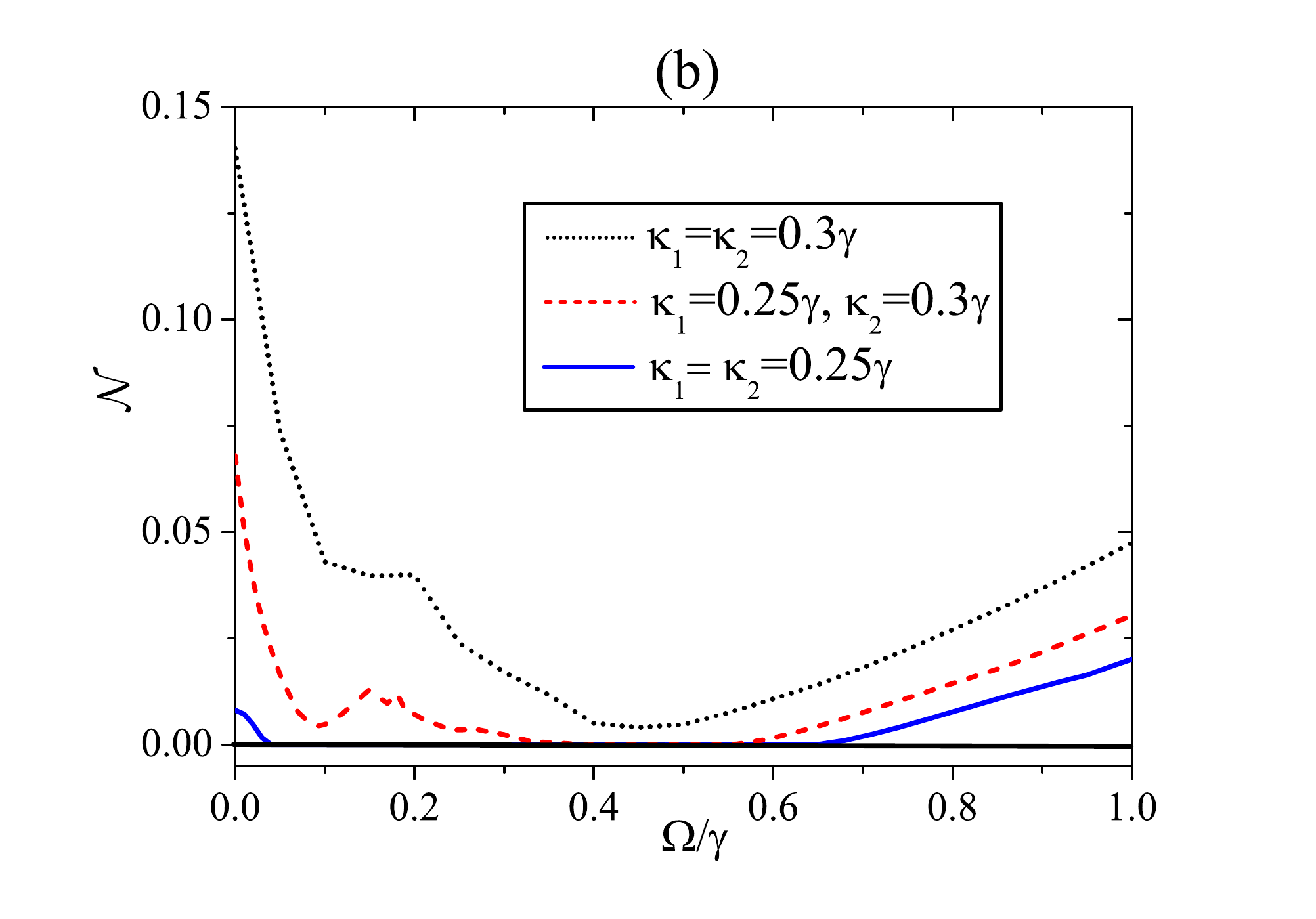}}
\end{center}
\caption{The non-Markovianity $\mathcal{N}$ as a function of the coupling
constant $\Omega /\gamma$ between two modes that are dissipated by
Lorentzian reservoirs with $\gamma_{1}=\gamma_{2}=\gamma$.
(a) and (b) demonstrate weak and strong couplings between
the qubit and the modes in terms of $\kappa_{1}$ and $\kappa_{2}$.
The other parameters are chosen as $\lambda_{1}=\lambda_{2}=0.5\gamma$
in (a) and $\lambda_{1}=\lambda_{2}=0.8\gamma$ in (b).}
\label{NM-2}
\end{figure}

We again assume the two modes and the corresponding reservoirs are initially in their ground states with 
only up to one excitation present in the total system. The qubit can initially be in a (normalized) superposition $\ket{\psi}_s=c_{0}(0)\ket{0}_s+h(0)\ket{1}_s$, so that the initial total state is 
$\ket{\Phi(0)}=\ket{\psi}_s\ket{00}_{m_1m_2}\ket{\overline{\mathbf{0}}\overline{\mathbf{0}}}_{\mathcal{R}_{1}\mathcal{R}_{2}}$ with
$\left|\overline{\mathbf{0}}\right\rangle_{\mathcal{R}_{n}}\equiv\prod_{k}\left|0_{k}\right\rangle_{\mathcal{R}_{n}}$. 
The total evolved pure state then reads
\begin{eqnarray}\label{state-t}
\left|\Phi(t)\right\rangle&=&\left[c_{0}(0)\left|0\right\rangle_{s}
+h(t)\left|1\right\rangle_{s}\right]\left|00\right\rangle_{m_{1}m_{2}}
\left|\overline{\mathbf{0}}\overline{\mathbf{0}}\right\rangle_{\mathcal{R}_{1}\mathcal{R}_{2}}\nonumber\\
&+&\left|0\right\rangle_{s}[c_{1}(t)\left|10\right\rangle_{m_{1}m_{2}}
+c_{2}(t)\left|01\right\rangle_{m_{1}m_{2}}]\left|\overline{\mathbf{0}}\overline{\mathbf{0}}\right\rangle_{\mathcal{R}_{1}\mathcal{R}_{2}}\nonumber\\
&+&\left|0\right\rangle_{s}\left|00\right\rangle_{m_{1}m_{2}}
\sum_{n=1}^2 \sum_{k}c_{n,k}(t)\left|\mathbf{1}_{k}\right\rangle_{\mathcal{R}_{n}}\ket{\overline{\mathbf{0}}}_{\mathcal{R}_{\bar{n}}},
\end{eqnarray}
where $\ket{\mathbf{1}_k}_{\mathcal{R}_{n}}\equiv \ket{0\cdots1_k\cdots0}_{\mathcal{R}_{n}}$ means that there is one excitation in the $k$th mode
of the reservoir $\mathcal{R}_{n}$ and $\bar{n}$ is the complementary of $n$ (i.e., $\bar{n}=2$ if $n=1$ and viceversa). 
The initial conditions of the coefficients appearing in $\ket{\Phi(t)}$ are $c_1(0)=c_2(0)=c_{n,k}(0)=0$.
From the Schr\"{o}dinger equation \cite{open}, the time evolution of the total system
in the interaction picture with the Hamiltonian of Eq.~(\ref{H2}) is determined by the following differential equations
\begin{eqnarray} \label{coefft}
\dot{h}(t) &=& -i\kappa_{1}c_{1}(t)-i\kappa_{2}c_{2}(t), \nonumber\\
\dot{c}_{1}(t) &=&-i\kappa_{1}h(t)-i\Omega c_{2}(t)-ig_{1,k}^{*}e^{i \Delta_{1,k} t}c_{1,k}(t),\nonumber\\
\dot{c}_{2}(t) &=&-i\kappa_{2}h(t)-i\Omega c_{1}(t)-ig_{2,k}^{*}e^{i \Delta_{2,k} t}c_{2,k}(t), \nonumber\\
\dot{c}_{1,k}(t) &=& -ig_{1,k}^{*}e^{i \Delta_{1,k} t}c_{1}(t), \nonumber\\
\dot{c}_{2,k}(t) &=& -ig_{2,k}^{*}e^{i \Delta_{n,k} t}c_{2}(t).
\end{eqnarray}
Integrating the last two equations with the initial condition $c_{n,k}(0)=0$ ($n=1,2$) and inserting their solutions into the second and third 
equation above, one obtains two integro-differential equations for the amplitudes $c_{1}(t)$
and $c_{2}(t)$
\begin{eqnarray}\label{c1c2}
\dot{c}_{1}(t)&=&-i\kappa_{1}h(t)-i\Omega c_{2}(t)\nonumber\\
&&-\int_{0}^{t}\sum_{k}|g_{1,k}|^{2}e^{-i\Delta_{1,k}(t-t^{\prime})}c_{1}(t^{\prime})dt^{\prime},\nonumber\\
\dot{c}_{2}(t)&=&-i\kappa_{2}h(t)-i\Omega c_{1}(t)\nonumber\\
&&-\int_{0}^{t}\sum_{k}|g_{2,k}|^{2}e^{-i\Delta_{2,k}(t-t^{\prime})}c_{2}(t^{\prime})dt^{\prime}.
\end{eqnarray}
The sum $\sum_{k}|g_{n,k}|^{2}e^{i(\omega_{0}-\omega_{n,k})(t-t^{\prime})}$ in the above equations is
recognized as the correlation function $f_{n}(t-t^{\prime})$ of the reservoir $\mathcal{R}_{n}$, which in the limit of a large number of modes can be changed into an integration in terms of the spectral density $J_{n}(\omega)$ as \cite{open}
\begin{equation}
f_{n}(t-t^{\prime})=\int d\omega J_{n}(\omega)\exp [i(\omega _{0}-\omega )(t-t^{\prime })].
\label{Fourier}
\end{equation}
We take each reservoir $\mathcal{R}_{n}$ with a Lorentzian spectral density 
$J_{n}(\omega)=\gamma _{n}\lambda _{n}^{2}/\{2\pi [(\omega -\omega _{0})^{2}+\lambda_{n}^{2}]\},$ 
where $\gamma _{n}$ is the mode-reservoir coupling strength
and $\lambda _{n}^{-1}$ the reservoir correlation time \cite{open,lofrancoreview}. 
The two-point correlation function of Eq.~(\ref{Fourier}) can be then expressed as $f_{n}(\tau )=\frac{1}{2}\gamma _{n}\lambda _{n}\exp (-\lambda _{n}|\tau |)$.
Therefore, the amplitudes $h(t)$, $c_{1}(t)$ and $c_{2}(t)$ can be obtained by solving the first one of Eqs. (\ref{coefft}) together with Eq.~(\ref{c1c2}) by using the standard Laplace transform technique. The reduced dynamics of the qubit and the of other parts of the overall system are then determined by tracing out the opportune degrees of freedom from the evolved total state $\ket{\Phi(t)}$ of Eq.~(\ref{state-t}).

In Fig.~\ref{NM-2}, we plot the non-Markovianity $\mathcal{N}$ as a function of
the scaled mode-mode coupling strength $\Omega/\gamma$ for different values of
system-mode couplings $\kappa_{1}$, $\kappa_{2}$ and assuming $\gamma_{1}$=$\gamma_{2}$=$\gamma$.
As shown in Fig.~\ref{NM-2}(a), when the memory effects of the two reservoirs alone (that is with $\Omega/\gamma=0$) are not sufficient to make the system experience non-Markovian dynamics, the introduction of mode-mode coupling with sufficient strength
can drive the Markovian dynamics to the non-Markovian one. The non-Markovianity
is moreover proportional to the coupling strength for given values of $\kappa_{n}$ and $\lambda_{n}$.
This implies that the coupling of the modes, as constituents of the compound memory of the
overall environment, can further enhance the memory effects of the latter on the qubit.
On the other hand, if the system already undergoes non-Markovian dynamics without mode-mode coupling,
the relation between the non-Markovianity and the mode-mode coupling results to be nonmonotonic.
The mode-mode coupling reduces the non-Markovianity and can even transform a non-Markovian dynamics ($\mathcal{N}>0$) 
into a Markovian one ($\mathcal{N}=0$). Nevertheless, the further increase
of the coupling strength $\Omega$ can recover and increase the non-Markovianity, as shown in Fig.~\ref{NM-2}(b).
These behaviors are analogous to the ones found before for the case of memoryless reservoirs, a slight difference being that here the non-Markovianity exhibits oscillations as $\Omega/\gamma$ increases before reaching its minimal value (see Figs.~\ref{NM}(b) and \ref{NM-2}(b)).

\section{Conclusion}\label{conclusion}
In this paper we have addressed the study of the effects of the coupling between two parts of a multiple environment on the dynamics of a quantum system. In particular, we have considered a qubit (the quantum memory) simultaneously interacting with two coupled bosonic modes (the control devices) which are in turns dissipated into memoryless or memory-keeping reservoirs. 
In the case of memoryless reservoirs, we have proven that the two cavity modes play the role of unique memory sources of the
overall environment for the qubit and their coupling can be thus viewed as a coupling between two quantum memory sources. 
In the case of memory-keeping reservoirs, the two modes are instead
the constituent parts of the total memory source of the overall environment and their coupling can be now meant as a coupling between two partial memory sources for the qubit. 
We have shown that in both cases the Markovian dynamics of the qubit, existing without the mode-mode coupling (the control parameter),
can become non-Markovian by adjusting the control parameter over a certain threshold.
Moreover, higher values of the control parameter enable larger non-Markovianity for the qubit. 
Differently, when the qubit evolution is already in a non-Markovian regime for a zero mode-mode coupling, a nonmonotonic relationship
arises between non-Markovianity and control parameter. 
Namely, multiple crossovers from non-Markovian to Markovian regimes may occur by increasing the mode-mode coupling.
This may appear surprising since, on the basis that increasing the coupling between memory sources for the qubit entails a transition from Markovian to non-Markovian regimes for the qubit, one expects that an increasing of the control parameter always induces an enhancement of memory effects on the qubit dynamics.

We remark that the behaviors above happen independently of the nature of the reservoirs. Our findings evidence that when the environment is composite the underlying physical mechanisms may be counterintuitive.    
The environmental coupling thus reveals as a powerful and effective tool to activate and harness quantum non-Markovianity of open systems. It is worth to notice that our system has the advantage to make it emerge in a clear way the effects of this coupling on the dynamics of a quantum system and, at the same time, to be simple enough to find feasibility within current experimental technologies, for instance in circuit QED \cite{circuitQED} or in simulating all-optical setups \cite{chiuriSciRep}.
Since non-Markovianity is linked to a dynamical recovery of the quantum coherence of a qubit \cite{open,RivasReview}, our work highlights that engineering and exploiting suitably structured compound environments can supply useful developments for controlling and preserving quantum memory resources. It also motivates further studies regarding the effects of multiple environments on the dynamics of correlations in many-qubit systems.

\begin{acknowledgments}
In this work Z.X.M. and Y.J.X. are supported by the National Natural Science Foundation (China) under Grants Nos. 11204156, 61178012 and 11247240,
the Promotive Research Fund for Excellent Young and Middle-Aged Scientists of Shandong Province (China) under Project No. BS2013DX034, and the Open Project of Key Laboratory of Quantum Information (CAS) under Grant No. KQI201503. 
R.L.F. acknowledges support by the Brazilian funding agency CAPES [Pesquisador Visitante Especial - Grant No. 108/2012].
\end{acknowledgments}

\end{document}